\documentclass[epj,nopacs]{svjour}
\usepackage{latexsym}
\usepackage{graphics}

\begin{document}

\thispagestyle{empty}
\title{
Thermal Casimir effect in ideal metal rectangular boxes
}

\author{
B.~Geyer \and
G.~L.~Klimchitskaya\thanks{on leave from 
North-West Technical University, St.Petersburg, Russia}
\and V.~M.~Mostepanenko\thanks{on leave from Noncommercial Partnership
``Scientific {\protect \\} Instruments'',  Moscow,  Russia}
}

\institute{
Center of Theoretical Studies and Institute for Theoretical
Physics, Leipzig University, {\protect \\}
Vor dem Hospitaltore 1, 100920,
D-04009, Leipzig, Germany 
}
\date{Received: date / Revised version: date}
\abstract{
The thermal Casimir effect in ideal metal rectangular 
boxes is considered using the method of zeta functional 
regularization. The renormalization procedure is suggested 
which provides the finite expression for the Casimir free 
energy in any restricted quantization volume. This 
expression satisfies the classical limit at high 
temperature and leads to zero thermal Casimir force for 
systems with infinite characteristic dimensions. In the 
case of two parallel ideal metal planes the results, as 
derived previously using thermal quantum field theory in 
Matsubara formulation and other methods, are reproduced 
starting from the obtained expression. It is shown that 
for rectangular boxes the temperature-dependent 
contribution to the electromagnetic Casimir force can be 
both positive and negative depending on side lengths. 
The numerical computations of the scalar and 
electromagnetic Casimir free energy and force are 
performed for cubes.
}
%\pacs{11.10.Wx, 03.70.+k, 11.10.Gh}
\authorrunning{ B.~Geyer et al.}
\titlerunning{Thermal Casimir effect in ideal metal rectangular boxes}

\maketitle

\section{Introduction}
\label{intro}
During the last few years there was an increasing interest to the Casimir
effect \cite{1}, the physical phenomenon which is caused by the
modification of the spectrum of zero-point and thermal oscillations
in restricted quantization volumes and in spaces with nontrivial
topology \cite{2,3}. This effect has found prospective fundamental
applications, e.g., for constraining predictions of modern unification
theories beyond the standard model \cite{4,5,6,7,8,8a} or in the braneworld
cosmological scenaria \cite{9,9a,9b}. Although much of recent research on
the Casimir effect was concentrated in the experimental issues and
material properties \cite{10,11,12}, some of the principal
field-theoretical aspects remain unsettled until the present time.

One of the most intriguing results is that
the Casimir energy and force
may change sign depending on geometry of the configuration and the type of
boundary conditions. A dramatic example of this situation
which has given rise to many discussions in the literature for several
decades is a rectangular box with side lengths $a,\,b$ and $c$.
Lukosz \cite{13} noticed that the electromagnetic Casimir energy
inside an ideal metal box  may change sign
depending on side sizes $a,\,b$ and $c$.
The detailed investigation of the Casimir energy for
fields of different spins, when it is positive or negative, inside
a rectangular box as a function of box dimensions was performed by
Mamayev and Trunov \cite{14,15}. In particular, the analytical
results for two- and three-dimensional boxes 
at zero temperature were obtained by the
repeated application of the Abel-Plana formula \cite{2,10}.
Ambj{\o}rn and Wolfram \cite{16} used the Epstein zeta function
to calculate the Casimir energy for a scalar and electromagnetic
field in hypercuboidal regions in $n$-dimensional space-time.
The electromagnetic Casimir densities were investigated for
a wedge \cite{16a} and for a wedge with a coaxial cylindrical
shell \cite{16b}.
The problem of isolation of the divergent terms in the vacuum energy
and their interpretation received the most attention.
In recent years this problem was reformulated as
 a rectangular box or arbitrary shaped cavity
divided into two sections by an ideal metal movable partition
(piston) \cite{17,18,19,20,21,22}.
It was shown that the Casimir force acting on the piston
with Dirichlet boundary conditions
attracts it to the nearest wall.
Based on this, some doubts about the previously
obtained results demonstrating Casimir repulsion in cubes have
been raised.
Both sets of results obtained for standard boxes and for boxes with 
a piston are, however, in mutual agreement. The attraction (or repulsion for
a piston with Neumann boundary conditions \cite{22}) of a piston to the
nearest face of the box with any sides $a,\,b,\,c$ does not negate the
Casimir repulsion for boxes without a piston that have some appropriate
ratio of $a,\,b$ and $c$. The point is that the cases with
an empty space outside the box and with another section
of the larger box ouside the piston are quite different. In the first case
the vacuum energy outside the box does not depend on  $a,\,b$ and $c$
and there is no force acting on the box from the outside.
Whereas in the second case there is an extra section of the larger box
outside the piston which gives rise to the additional force acting on it.

Another intriguing problem to which this paper is devoted is the problem
of the thermal Casimir effect in ideal metal rectangular boxes.
The first calculations on this subject \cite{16} resulted in a divergent
free energy after removing the regularization. More recent results appear
to be either infinite \cite{23} or ambiguous \cite{24}. Reference \cite{25}
reconsidered the derivation of the Casimir free energy for a massless scalar
and electromagnetic field using zeta functional regularization. However,
as shown below, the used formalism does not include all necessary
subtractions. This led to unjustified conclusions on the behavior of the
Casimir free energy as a function of temperature. Specifically, it is
claimed that in rectangular cavities in the high temperature regime the
leading term of the free energy is
of order $(k_BT)^4$, where $k_B$ is the Boltzmann constant \cite{25}.
The terms of orders $(k_BT)^3$ and $(k_BT)^2$ are also obtained \cite{25}.
The coefficients near such terms unavoidably depend of $a,\>b$ and $c$ and,
thus, they contribute to the Casimir force.

In the present paper
we consider the thermal Casimir effect in rectangular boxes starting
from the general expression for the Casimir free energy in the framework
of Matsubara quantum field theory at nonzero temperature. 
The thermal correction in this expression is renormalized in the same
way as the zero-temperature contribution.
We critically reconsider the known results from a unified point of
view, separating clearly the empty space contribution
in the volume of a body and other geometrical terms not only in
the Casimir energy at zero temperature, but also in the temperature
correction to it.
Using the method of zeta functional regularization \cite{25a,26a,26b}, 
the finite expression
for the Casimir free energy associated with a volume $V$ is obtained. This
expression is first applied to the familiar case of two plane parallel
planes and the standard results \cite{2,3,10,26} are reproduced. 
The Casimir force, as defined in this paper, goes to zero when the 
characteristic dimensions of the volume $V$ go to infinity. This is achieved
by a subtraction of the free energy of the black body radiation within the 
volume $V$ and of two other geometrical contributions of quantum origin. 
Then we apply 
obtained general expression for the Casimir free energy
to the cases of the massless scalar field with Dirichlet boundary conditions
and the electromagnetic field in rectangular boxes.
We show that the temperature-dependent contribution to the Casimir force
at zero temperature can be both positive and negative (i.e., it leads to
repulsion or attraction) depending on the box side lengths.
This property is preserved for boxes with zero electromagnetic Casimir
force at zero temperature. Numerical computations of the scalar and 
electromagnetic Casimir force as functions of side length and temperature
for cubes are performed.

The paper is organized as follows. In Sect.~2 the general expression for the
Casimir free energy is derived using the method of zeta functional
regularization. Section~3 contains the application of this expression
to the case of two plane parallel planes. In Sects.~4 and 5 the massless
scalar and the electromagnetic fields, respectively, are considered
in 3-dimensional rectangular boxes. Section~6 contains our conclusions and
discussion. Throughout the paper the system of units is used where
$c=\hbar=1$.

\section{The Casimir energy in Matsubara formulation and zeta
functional regularization}

In the Matsubara formalism one uses an Euclidean field theory, considered
as a continuation of the field theory in Minkowski space-time by 
a rotation of the time
$t\to -{\rm i} \tau$. The Euclidean time $\tau$   is confined to the
interval $\tau\in[0,\beta]$, where $\beta=1/(k_BT)$ is
the equivalent dimension corresponding to the inverse temperature.
The  bosonic
fields must be periodic,
$\varphi(\tau+\beta,\mbox{\boldmath$r$})=
\varphi(\tau,\mbox{\boldmath$r$})$.
In the limit of zero temperature one re-obtains the theory
on the whole time axis.

In the Matsubara formalism, the partition function $Z$ 
has the following representation in terms of a functional integral,
\begin{equation}
Z=C\int D\varphi\, e^{-S_E[\varphi]},
\label{5Z1}
\end{equation}
where $S_E[\varphi]$ is the Euclidean action. It can be obtained from
the corresponding one in Minkowski space-time  by the
replacement $S$ with ${\rm i}S_E$. For example,
for a scalar field  with mass $m$ it is
\begin{equation}
S_E[\phi]=\frac12 \int_0^\beta d\tau \,\int d\mbox{\boldmath$r$}\
\varphi K_E \, \varphi,
\label{5SE}\end{equation}
where
\begin{equation}
K_E=\left(-\Box_E+m^2\right).
\label{5KE}
\end{equation}
The Euclidean wave operator
$\Box_E=\partial^2/\partial \tau^2+\Delta$
is the continuation of the D'Alembert operator.  In the functional
integral (\ref{5Z1}), the field to be integrated over must fulfill
the corresponding periodicity conditions.

In general, in the Matsubara formalism, the construction of the theory,
to a large extent, goes in parallel to the zero temperature case.
 This is true for instance
in the calculation of the functional integral.  Since we 
consider free field theories, the functional integral is Gaussian
and can be calculated directly. Using the
infinite dimensional analogue to the standard finite dimensional
relations, we obtain for the
partition function from (\ref{5Z1})
\begin{equation}
Z=C\left(\det K_E\right)^{-1/2},
\label{5Z2}\end{equation}
where $C$ is an irrelevant constant which will be dropped
below. Further, for the free energy we get
\begin{equation}
{\cal F}=-\frac{1}{\beta}\ln Z=\frac{1}{2\beta}\mbox{Tr} \ln K_E.
\label{5F1}\end{equation}
The trace in this expression is taken over the same space of fields to
be integrated over in the functional integral (\ref{5Z1}).

Since we assume thermal equilibrium, it is always possible to
separate the Euclidean time variable from the spatial variables.
Assuming for the spatial part an eigenfunction expansion 
\begin{equation}
-\mbox{\boldmath$\nabla$}^2 \Phi_J(\mbox{\boldmath$r$})=
\Lambda_J \Phi_J(\mbox{\boldmath$r$}), \quad
\Lambda_J\equiv\omega_J^2-m^2,
\label{5EF}\end{equation}
where $J$ is a collective index,
we obtain a basis in the space of fields $\varphi$,
\begin{equation}
\Phi_{l,J}(\tau,\mbox{\boldmath$r$})=
\frac{{\rm e}^{-{\rm i}\xi_l \tau}}{\sqrt{2\pi}}\,  
\Phi_J(\mbox{\boldmath$r$}),
\label{5EF1}
\end{equation}
with the so-called Matsubara frequencies
\begin{equation}
\xi_l=2\pi k_BTl, \qquad l=0,\,\pm 1,\,\pm 2,\,\ldots\,.
\label{5oml}
\end{equation}
These functions are periodic in $\tau$  and these are eigenfunctions of $K_E$,
\begin{equation}
K_E \Phi_{l,J}=\left(\xi_l^2+\Lambda_J+m^2\right)\Phi_{l,J}.
\label{5EF2}
\end{equation}
As a consequence, the trace in the free energy becomes a sum over
the logarithm of  the eigen\-values,
\begin{equation}
{{\cal F}}_0=\frac12 k_BT\sum_{l=-\infty}^{\infty}
\sum_J\ln\left(\xi_l^2+\Lambda_J+m^2\right).
\label{5F2}
\end{equation}
Note that if the field is massless and all quantum numbers in the
collective index $J$ are discrete, it is assumed in 
(\ref{5F2}) that there are no physical states
with $\Lambda_J=0$.

 In (\ref{5F2}), ${\cal F}_0$ is the 
free energy  of an ensemble of  states
containing particles at the temperature $T$.
In the special case of $T\to0$, the time interval stretches over
the whole axis and the sum over the Matsubara frequencies  becomes
the integral over the frequency $\xi$,
\begin{equation}
k_BT\sum_{l=-\infty}^{\infty} f(\xi_l)\to \int_{-\infty}^\infty
\frac{d\xi}{2\pi}f(\xi)
\label{5si}
\end{equation}
[here $f(\xi_l)$ is a function which must allow for an analytic
continuation from discrete values to continuous ones]. In this way,
${{\cal F}}_0$ defined in  (\ref{5F2}) turns into the vacuum energy
$E_{0}$.

The free energy, as given by (\ref{5F2}), still contains ultraviolet
divergences and one has to introduce a regularization. 
In zeta functional regularization, the free energy becomes
\begin{equation}
{{\cal F}}_0(s)=-\frac12\frac{\partial}{\partial s}\mu^{2s}
k_BT\sum_{l=-\infty}^{\infty}
\sum_J\left(\xi_l^2+\Lambda_J+m^2\right)^{-s}.
\label{5F3}
\end{equation}
The regularization is removed for $s\to0$ and $\mu$ is an arbitrary
parameter with the dimension of mass. The separation of the ultraviolet
divergences can be done quite easily because these are the same as at
zero temperature. There are two ways to proceed.
In the first method one has to apply
the Abel-Plana formula \cite{2,10} to the frequency sum in (\ref{5F3}).
This way has the advantage that it can also be applied to the case when
there is an additional dependence on $l$, for example through the
dielectric permittivity
entering in the form of $\xi^2\to \varepsilon({\rm i}\xi)\xi^2$.
Another, to some extent easier, way is through the application of the
Poisson summation formula \cite{27}. According to
this formula, if $c(\alpha)$ is the Fourier transform of a function
$b(x)$,
\begin{equation}
c(\alpha)=\frac{1}{2\pi}\int_{-\infty}^{\infty}b(x)
{\rm e}^{-{\rm i}\alpha x}dx,
\label{5Poi1}
\end{equation}
\noindent
then it follows that
\begin{equation}
\sum_{l=-\infty}^{\infty}b(l)=2\pi\sum_{l=-\infty}^{\infty}c(2\pi l).
\label{5Poi2}
\end{equation}

By putting
\begin{equation}
b(x)={\rm e}^{-zx^2}, \qquad
c(\alpha)=\frac{1}{2\sqrt{\pi z}}{\rm e}^{-\frac{\alpha^2}{4z}},
\label{5ex}
\end{equation}
\noindent
we obtain from (\ref{5Poi2}) the following equality:
\begin{equation}
\sum_{l=-\infty}^{\infty}{\rm e}^{-zl^2}=\sqrt{\frac{\pi}{z}}
\sum_{n=-\infty}^{\infty}{\rm e}^{-\pi^2n^2/z},
\label{Poisson}\end{equation}
where ${\rm Re}\, z>0$ is assumed.
In order to use this equality we represent (\ref{5F3}) 
as a parametric integral,
\begin{equation}
{{\cal F}}_0(s)=-\frac{1}{2\beta}\frac{\partial}{\partial s}\mu^{2s}
\int_0^\infty\frac{dt}{t}\frac{t^s}{\Gamma(s)}
\sum_{l=-\infty}^{\infty}\sum_J
{\rm e}^{-t\left(\xi_l^2+\Lambda_J+m^2\right)}
\label{5F4}
\end{equation}
and apply (\ref{Poisson}) 
with $z=(2\pi k_BT)^2t=(2\pi/\beta)^2t$.
The result is
\begin{eqnarray}
&&
{{\cal F}}_0(s)=-\frac12\frac{\partial}{\partial s}\mu^{2s}
\sum_{n=-\infty}^{\infty}
\int_0^\infty\frac{dt}{t}\frac{t^s}{\Gamma(s)\sqrt{4\pi t}}
\nonumber \\
&&~~~~~~~~~~~~~~~~~~~~
\times \sum_J
{\rm e}^{-\frac{n^2\beta^2}{4t}-t\left(\Lambda_J+m^2\right)}  .
\label{5F4a}
\end{eqnarray}

The $n$-dependent factor in the exponential provides convergence for
the $t$-inte\-gra\-tion at $t\to0$ for all terms in the sum over
$n$ except for $n=0$. The latter is just the zero temperature contribution.
This can be seen by applying (\ref{5si}) to the frequency sum in 
(\ref{5F4}),
\begin{equation}
k_BT\sum_{l=-\infty}^{\infty}{\rm e}^{-t \xi_l^2}
\raisebox{-3pt}{$\to\atop \scriptscriptstyle T\to0$}
\int_{-\infty}^\infty\frac{d\xi}{2\pi}{\rm e}^{-t\xi^2 } =
\frac{1}{\sqrt{4\pi t}}.
\label{5e24}
\end{equation}
As a consequence, we can split the free energy
into the zero temperature part and the temperature dependent
addition $\Delta_T{\cal F}_0$ as
\begin{equation}
{\cal F}_0(s)=E_{0,{\rm eff}}(s)+\Delta_T{{\cal F}}_0(s).
\label{5F0T}
\end{equation}
Here, the vacuum energy at zero temperature is
\begin{equation}
E_{0,\rm eff}(s)    =
    -\frac12\frac{\partial}{\partial s}\mu^{2s}
    \int_{-\infty}^\infty\frac{d\xi}{2\pi}
        \sum_J \left(\xi^2+\Lambda_J+m^2\right)^{-s}.
\label{5F5}
\end{equation}
The nonrenormalized thermal correction is given by
\begin{eqnarray}
&&
\Delta_T{{\cal F}}_0(s)=-\frac{\partial}{\partial s}\mu^{2s}
\sum_{n=1}^{\infty}
 \int_0^\infty\frac{dt}{t}\frac{t^s}{\Gamma(s)\sqrt{4\pi t}}
\nonumber \\
&&~~~~~~~~~~~~~~~~~~~~
\times  \sum_J
{\rm e}^{-\frac{n^2\beta^2}{4t}-t\left(\Lambda_J+m^2\right)}  .
\label{5e23a}
\end{eqnarray}
\noindent
Note that (\ref{5F0T}) has a transparent physical interpretation
only for the temperature independent
boundary conditions considered here.

The ultraviolet divergences are contained in $E_{0,\rm eff}(s) $ and can
be dealt with in the standard way \cite{2,10}.
This results in the replacement of $E_{0,\rm eff}(s)$ with
$E_0^{\rm ren}$ in (\ref{5F0T}), where $E_0^{\rm ren}$ is the finite
renormalized Casimir energy at zero temperature associated with a
volume $V$, which goes to zero when $V\to\infty$.
In $\Delta_T{\cal F}_0(s)$ in (\ref{5e23a}), the integration over $t$ is
convergent and we can remove the regularization, i.e., we can put $s=0$
using the equality
\begin{equation}
\lim_{s\to 0}\frac{\partial}{\partial s}\,\frac{f(s)}{\Gamma(s)}=f(0),
\label{5e23b}
\end{equation}\noindent
where $f(s)$ is any regular function at $s=0$.
Following this, the integration over $t$  and the summation over $n$
can be carried out explicitly
\begin{equation}
\Delta_T{\cal F}_0=k_BT\sum_J\ln\left(1-e^{-\beta\sqrt{\Lambda_J+m^2}}\right).
\label{5F7}
\end{equation}
In this formula, we see the sum of the $T$-dependent contributions to
the free energies of the individual degrees of freedom, or modes
$\Lambda_J$, of the considered system.
Taking into account $\Lambda_J+m^2=\omega_J^2$, the total free
energy of all the oscillator modes appears to be
\begin{equation}
{\mathcal F}_0=E_0^{\rm ren}+k_BT\sum_{J}\ln\bigl(1-
{\rm e}^{-\beta\omega_J}\bigr),
\label{5e24a}
\end{equation}
\noindent
where  the zero temperature contribution is already
replaced with $E_0^{\rm ren}$.
For instance, if we take the volume $V$ to be that of empty space,
the modes are plane waves, the index $J$ becomes the wave vector
$\mbox{\boldmath$k$}$ and the sum over $J$ turns into
the corresponding momentum
integration with respect to $d\mbox{\boldmath$k$}/(2\pi)^3$.
As a result, from (\ref{5F7}) we obtain
 the free energy density
of the blackbody radiation
\begin{equation}
f_{\rm bb}(T)=k_BT\int\frac{d\mbox{\boldmath$k$}}{(2\pi)^3}
    \ln\left(1-
e^{-\beta|\mbox{\scriptsize\boldmath$k$}|}\right)
    =-\frac{\pi^2 (k_BT)^4}{90}.
\label{5bb}
\end{equation}
We note that (\ref{5bb}) holds for a scalar field. For the
electromagnetic field one would have to multiply by a factor of 2.
In fact, for the empty space $f_{\rm bb}$ defined in (\ref{5bb})
is the complete free energy. This is because in this case the zero
temperature part was the vacuum energy of empty space which was
disregarded in the replacement of $E_{0,{\rm eff}}$ with
$E_0^{\rm ren}$. Using the thermodynamic connection between the
energy at a temperature $T$ and the free energy
\begin{equation}
U(T)=-T^2\frac{\partial}{\partial T}\,\frac{{\cal F}(T)}{T},
\label{5e31a}
\end{equation}
\noindent
it is evident that for
${\cal F}(T)=f_{\rm bb}^{\rm em}(T)=2f_{\rm bb}(T)$ the
respective energy density is in agreement with Planck's
blackbody radiation density
\begin{equation}
u=\frac{\pi^2(k_BT)^4}{15}.
\label{5e31b}
\end{equation}

If we consider the free energy in a
restricted volume $V$, then in accordance with (\ref{5e24a}),
we have to keep the zero temperature part $E_0^{\rm ren}$, where the
empty space contribution has been dropped already. 
For  the temperature dependent part,
we have to note that we are interested in the change in energy
which comes from the volume $V$.
Therefore we need to subtract from the temperature dependent part
$\Delta_T{\cal F}_0$ of the free energy, the corresponding amount related to
empty space, i.e., the blackbody radiation density $f_{\rm bb}$
multiplied by the volume $V$. As a result, we come to the following
expression for the renormalized free energy associated with a finite
volume $V$,
\begin{equation}
{\cal F}=E_0^{\rm ren}+\Delta_T{\cal F}_0-Vf_{\rm bb},
\label{5Ff}
\end{equation}
which can be considered as the relevant quantity for 
the Casimir effect at finite temperature. In the next section we show 
that this is the case for the configuration of two parallel ideal metal
planes.

However, in general case (\ref{5Ff}) is not satisfactory and requires
at least two additional subtractions in order to get the physical
Casimir free energy. 
The point is that the asymptotic expression of
the thermal correction, $\Delta_T{\cal F}_0$, in the limit of
high temperatures (large separations)  
in addition  to the leading term,
\begin{equation}
-V\frac{\pi^2(k_BT)^4}{90(\hbar c)^3},
\label{eq29a}
\end{equation}
\noindent
contains two other similar terms
\begin{equation}
\alpha_1\frac{(k_BT)^3}{(\hbar c)^2}+\alpha_2\frac{(k_BT)^2}{\hbar c},
\label{eq29b}
\end{equation}
\noindent
where $\alpha_1$ and $\alpha_2$ have dimensions of cm${}^2$ and cm,
respectively. They are expressed through the  heat kernel coefficients
$a_{1/2}$ and $a_1$, respectively
\cite{10}. This was shown quite generally by Dowker and Kennedy \cite{28a}.
We use usual units in (\ref{eq29a}) and (\ref{eq29b}) to underline
the quantum character of all these terms. The coefficients
$\alpha_1$ and $\alpha_2$, generally speaking, depend on geometrical
parameters of the configuration (for two parallel planes 
$\alpha_1=\alpha_2=0$, but for a rectangular box $\alpha_1$ and $\alpha_2$
depend on the side sizes; see Sects.~4 and 5). Because of this, the
terms of the form (\ref{eq29b}) in the free energy would 
lead to forces of quantum
nature which   do not vanish
with increasing characteristic sizes 
of the body (the next expansion term in the
high-temperature limit of the free energy has the form of $\alpha_3k_BT$
with a dimensionless coefficient $\alpha_3$; it is of classical origin
and does not contribute to the Casimir force). Below we demonstrate
that for rectangular boxes the geometrical structure of the coefficients
$\alpha_1$ and $\alpha_2$ is just the same as in the two infinite terms
subtracted from the zero-temperature Casimir energy to make it finite
in the cutoff regularization \cite{2,10}.
Thus, one can perform additional, finite, renormalization of the
free energy resulting in
\begin{eqnarray}
&&
{\cal F}^{\rm phys}=E_0^{\rm ren}+\Delta_T{\cal F},
\label{eq29c}\\
&&
\Delta_T{\cal F}=
\Delta_T{\cal F}_0-Vf_{\rm bb}-
\alpha_1(k_BT)^3-\alpha_2(k_BT)^2.
\nonumber
\end{eqnarray}
\noindent
Here, the quantity $\Delta_T{\cal F}$ is the physical thermal
correction to the Casimir energy.
The respective Casimir force obviously vanishes when the
characteristic sizes of the volume $V$ along all three coordinate
axes go to infinity. 

Below we  use (\ref{eq29c}) to investigate the thermal
Casimir effect in  configurations  of two parallel ideal metal planes
(Sect.~3) and rectangular boxes (Sects.~4 and 5).
All other thermodynamic quantities like force and entropy can
be also derived from this formula.
As for configurations containing translational invariant directions
like parallel planes or a cylinder, one should bear in mind that all
quantities in (\ref{eq29c}) must be divided by corresponding parameters
like the area of  plates. 

\section{The electromagnetic Casimir free energy between two
parallel planes}

The Casimir free energy in the configuration of two parallel ideal metal
planes can be easily obtained from the general equation (\ref{eq29c}).
It is common knowledge that for two parallel planes the high-temperature
asymptotic expansion of $\Delta_T {\cal F}_0$ does not contain terms  
presented in  (\ref{eq29b}). Then
(\ref{eq29c}) and (\ref{5F7}) with $\alpha_1=\alpha_2=0$ represent
the electromagnetic Casimir free energy per unit area
in the form (prime adds a multiple 1/2 to the term with $n=0$)
\begin{eqnarray}
&&
{\cal F}_{\rm el}^{\rm phys}(a,T)=-\frac{\pi^2}{720a^3}
\label{7e53a}\\
&&
+
\frac{k_BT}{\pi}\int_{0}^{\infty}\!\!\!\!k_{\bot}dk_{\bot}
\sum_{n=0}^{\infty}{\vphantom{\sum}}^{\prime}
\ln\bigl(1-{\rm e}^{-\beta\omega_{k_{\bot},n}}\bigr)
+\frac{\pi^2(k_BT)^4a}{45},
\nonumber
\end{eqnarray}
\noindent
where
\begin{equation}
\omega_J=\omega_{k_{\bot},n}=\sqrt{k_{\bot}^2+\left(
\frac{\pi n}{a}\right)^2}.
\label{e53aa}
\end{equation}
\noindent
By introducing the new variable
\begin{equation}
z=\beta\omega_{k_{\bot},n}=\frac{1}{k_BT}
\sqrt{k_{\bot}^2+\Bigl(\frac{\pi n}{a}\Bigr)^2}
\label{7e53b}
\end{equation}
\noindent
and changing the order of summation and integration,
we rearrange (\ref{7e53a}) as
\begin{eqnarray}
&&
{\cal F}_{\rm el}^{\rm phys}(a,T)=-\frac{\pi^2}{720a^3}
\label{7e53c} \\
&&
+
\frac{(k_BT)^3}{\pi}
\sum_{n=0}^{\infty}{\vphantom{\sum}}^{\prime}
\int_{2\pi nt}^{\infty}\!\!\!z\,dz
\ln\bigl(1-{\rm e}^{-z}\bigr)
+\frac{\pi^2}{720a^3t^4},
\nonumber
\end{eqnarray}
\noindent
where $t\equiv T_{\rm eff}/T$, and the effective temperature is defined 
in the usual units as
$k_BT_{\rm eff}=\hbar c/(2a)$.
The integral entering (\ref{7e53c}) can be evaluated using the
series expansion
\begin{eqnarray}
&&
\int_{2\pi nt}^{\infty}\!\!\!z\,dz
\ln\bigl(1-{\rm e}^{-z}\bigr)=-\sum_{l=1}^{\infty}\frac{1}{l}
\int_{2\pi nt}^{\infty}\!\!\!z\,dz\,{\rm e}^{-lz}
\nonumber \\
&&~~~~~~~
=-\frac{1}{8\pi a^3}\sum_{l=1}^{\infty}\frac{1}{(lt)^3}
(1+2\pi nlt){\rm e}^{-2\pi nlt}.
\label{7e53d}
\end{eqnarray}
\noindent
Substituting this into (\ref{7e53c}) and performing the summation
in $n$, we obtain 
\begin{eqnarray}
&&
{\cal F}_{\rm el}^{\rm phys}(a,T)=-\frac{\pi^2}{720a^3}
\left\{1+\frac{45}{\pi^3}\sum_{l=1}^{\infty}\left[
\frac{\coth(\pi lt)}{t^3l^3}\right.\right.
\label{7e51} \\
&&~~~~~~~
\left.\left.
+\frac{\pi}{t^2l^2{\rm sinh}^2(\pi tl)}\right]-
\frac{1}{t^4}
\vphantom{\sum_{l=1}^{\infty}}
\right\}.
\nonumber
\end{eqnarray}
\noindent
 This is a well known result
\cite{2,3,10,26,28}, and the last term on the right-hand side of
(\ref{7e51}) originates from the subtracted contribution of
blackbody radiation. At low temperature $T\ll T_{\rm eff}$ (\ref{7e51})
leads to \cite{2,3,10,26,28}
\begin{equation}
{\cal F}_{\rm el}^{\rm phys}(a,T)=-\frac{\pi^2}{720a^3}
\left[1+\frac{45\zeta(3)}{\pi^3}
\left(\frac{T}{T_{\rm eff}}\right)^3-
\left(\frac{T}{T_{\rm eff}}\right)^4
\right],
\label{7e54}
\end{equation}
\noindent
where $\zeta(z)$ is the Riemann zeta function, and exponentially small
terms are omitted.
We emphasize that the quantity $1/(aT_{\rm eff})^3$ does not depend on
$a$. Because of this, only the last term on the right-hand side of
(\ref{7e54}), originating from the subtraction of the blackbody radiation
in (\ref{eq29c}),
contributes to the measurable quantity, the Casimir pressure following
from (\ref{7e54})
\begin{equation}
P_{\rm el}(a,T)=
-\frac{\partial{\cal F}_{\rm el}^{\rm phys}(a,T)}{\partial a}=
-\frac{\pi^2}{240a^4}
\left[1+\frac{1}{3}\,
\left(\frac{T}{T_{\rm eff}}\right)^4
\right].
\label{7e56}
\end{equation}
\noindent
Thus, if this subtraction would not be done in (\ref{eq29c}) and 
(\ref{7e53a}), only the exponentially small thermal corrections of
order $(T/T_{\rm eff})\exp({-2\pi T_{\rm eff}/T})$ were added to unity
on the right-hand side of (\ref{7e56}).

In a similar way, at high temperature $T\gg T_{\rm eff}$ (\ref{7e51})
results in \cite{2,3,10,26,28}
\begin{equation}
{\cal F}_{\rm el}^{\rm phys}(a,T)=
-\frac{k_BT}{8\pi a^2}\zeta(3),
\quad
P_{\rm el}(a,T)=
-\frac{k_BT}{4\pi a^3}\zeta(3).
\label{7e61}
\end{equation}
\noindent
This is the commonly known classical limit where the result does not depend
on the Planck constant \cite{29,30}. It is obtained because at high
temperature the free energy of the blackbody radiation is canceled
by the other terms contained in (\ref{7e51}). If the subtraction of the
blackbody radiation in (\ref{eq29c}) were not done, the term of order
$(k_BT)^4/(c\hbar)^3$ (in usual units)
would be dominating at high temperature (or large separations) in
contradiction with the classical limit.

\section{Scalar Casimir effect in rectangular boxes}

After the above demonstration that the general equation (\ref{eq29c}) leads
to the familiar result in the case of two parallel planes, we apply it
to calculate the thermal Casimir effect in rectangular boxes with sides
$a,\,b$ and $c$.
We start with the case of a massless scalar field with Dirichlet
boundary conditions. Then
(\ref{eq29c}), (\ref{5F7}) can be written in the form
\begin{eqnarray}
&&
{\cal F}^{\rm phys}(a,b,c,T)=E_0^{\rm ren}(a,b,c)
\nonumber \\
&&~~~
+k_BT\!\!
\sum_{n,l,p=1}^{\infty}\!\!\ln\bigl(1-{\rm e}^{-\beta\omega_{nlp}}
\bigr)
\nonumber \\
&&~~~
+\frac{\pi^2(k_BT)^4}{90}abc-\alpha_1(k_BT)^3-\alpha_2(k_BT)^2,
\label{8e86}
\end{eqnarray}
\noindent
where
\begin{equation}
\omega_{nlp}^2=\pi^2\left(\frac{n^2}{a^2}+\frac{l^2}{b^2}+
\frac{p^2}{c^2}\right)\!,\quad
n,\,l,\,p=1,\,2,\,3,\,\ldots
\label{8e86a}
\end{equation}
\noindent
and the Casimir energy at zero temperature is given by \cite{30a,31}
\begin{eqnarray}
&&
E_0^{\rm ren}(a,b,c)=
-\frac{\pi^2 bc}{1440a^3}+\frac{\zeta(3)(b+c)}{32\pi a^2}
-\frac{\pi}{96a}
\nonumber \\
&&~~
-\frac{\pi}{2a}\left[G\left(\frac{b}{a}\right)+
G\left(\frac{c}{a}\right)\right]-\frac{1}{a}
R\left(\frac{b}{a},\frac{c}{a}\right).
\label{8e42} 
\end{eqnarray}
\noindent
Here the following notations are introduced:
\begin{eqnarray}
&&
G(z)=-\frac{1}{2\pi}\sum_{n,l=1}^{\infty}
\frac{n}{l}K_1(2\pi nlz),
\nonumber \\
&&
R(z_1,z_2)\equiv\frac{z_1z_2}{8}\sum_{l,p=-\infty}^{\infty}
(1-\delta_{l0}\delta_{p0})
\label{8e11} \\
&&
~
\times\sum_{j=1}^{\infty}
\left(\frac{j}{\sqrt{l^2z_1^2+p^2z_2^2}}\right)^{3/2}
K_{3/2}\left(2\pi j\sqrt{l^2z_1^2+p^2z_2^2}\right)
\nonumber
\end{eqnarray}
\noindent
and $K_n(z)$ are Bessel functions of imaginary argument.

Equation (\ref{8e86}) presents the finite renormalized value
of the Casimir free energy for a scalar field with Dirichlet
boundary conditions in a box with sides
$a,\>b$ and $c$ valid at any temperature. 
However, the values of $\alpha_1$ and $\alpha_2$ remain unknown.
To determine them we should find the asymptotic expression for the 
nonrenormalized thermal
correction $\Delta_T{\cal F}_0$
at high temperatures (large separations). 
To do this, we identically rearrange $\Delta_T{\cal F}_0$
to the form
\begin{eqnarray}
&&
\Delta_T{\cal F}_0(a,b,c,T)=k_BTX(\beta_a,\beta_b,\beta_c),
\label{eq42a} \\
&&
X(\beta_a,\beta_b,\beta_c)\equiv\sum_{n,l,p=1}^{\infty}
\ln\left(1-{\rm e}^{-\sqrt{\beta_a^2n^2+\beta_b^2l^2+\beta_c^2p^2}}
\right),
\nonumber
\end{eqnarray}
\noindent
where
\begin{equation}
\beta_a=\frac{\pi\beta}{a}, \quad
\beta_b=\frac{\pi\beta}{b}, \quad
\beta_c=\frac{\pi\beta}{c}.
\label{eq42b}
\end{equation}
\noindent
Note that the quantity $X$ does not depend on $a,\>b,\>c$ and $T$
separately, but only through the products $aT,\>bT$ and $cT$.
Below we find the asymptotic expression of $X$ under the conditions
$\beta_a,\,\beta_b,\,\beta_c\ll 1$.
This can be done by the repeated application of the Abel-Plana
formula \cite{2,10}
\begin{equation}
\sum_{n=1}^{\infty}f(n)=-\frac{1}{2}f(0)+\int_{0}^{\infty}\!\!f(t)dt+
\int_{0}^{\infty}\!\!dt
\frac{f({\rm i}t)-f(-{\rm i}t)}{{\rm e}^{2\pi t}-1}.
\label{eq42c}
\end{equation}
\noindent
First we put
\begin{equation}
f(n)=\sum_{l,p=1}^{\infty}
\ln\left(1-{\rm e}^{-\sqrt{\beta_a^2n^2+\beta_b^2l^2+\beta_c^2p^2}}
\right).
\label{eq42d}
\end{equation}
\noindent
Then the application of (\ref{eq42c}) leads to
\begin{eqnarray}
&&
X(\beta_a,\beta_b,\beta_c)=-\frac{1}{2}\sum_{l,p=1}^{\infty}
\ln\left(1-{\rm e}^{-\sqrt{\beta_b^2l^2+\beta_c^2p^2}}
\right)
\nonumber \\
&&~~
+\int_{0}^{\infty}\!\!dt\sum_{l,p=1}^{\infty}
\ln\left(1-{\rm e}^{-\sqrt{\beta_a^2t^2+\beta_b^2l^2+\beta_c^2p^2}}
\right)
\nonumber \\
&&
+O(\ln\beta_a,\ln\beta_b,\ln\beta_c).
\label{eq42e}
\end{eqnarray}
\noindent
Here, it is taken into account that the last term on the right-hand side
of (\ref{eq42c}) with $f$ defined in (\ref{eq42d}) is
of order $\ln\beta_a,\>\ln\beta_b$ and $\ln\beta_c$.

Applying the Abel-Plana formula to each of the sums in (\ref{eq42e}),
we get
\begin{eqnarray}
&&
X(\beta_a,\beta_b,\beta_c)=\frac{1}{4}\sum_{p=1}^{\infty}
\ln\left(1-{\rm e}^{-\beta_cp}
\right)
\label{eq42g} \\
&&~~
-\frac{1}{2}\left(\frac{1}{\beta_a}+\frac{1}{\beta_b}\right)
\int_{0}^{\infty}\!\!dy\sum_{p=1}^{\infty}
\ln\left(1-{\rm e}^{-\sqrt{y^2+\beta_c^2p^2}}
\right)
\nonumber \\
&&~~
+\frac{1}{\beta_a\beta_b}\int_{0}^{\infty}\!\!dy
\int_{0}^{\infty}\!\!dv\sum_{p=1}^{\infty}
\ln\left(1-{\rm e}^{-\sqrt{y^2+v^2+\beta_c^2p^2}}
\right)
\nonumber \\
&&~~
+O(\ln\beta_a,\ln\beta_b,\ln\beta_c).
\nonumber
\end{eqnarray}
\noindent
Bearing in mind that
\begin{equation}
\sum_{p=1}^{\infty}\ln\left(1-{\rm e}^{-\beta_cp}\right)=
-\frac{\pi^2}{6\beta_c}+O(\ln\beta_c)
\label{eq42h}
\end{equation}
\noindent
and applying the Abel-Plana formula to the remaining two sums,
the following result is obtained:
\begin{eqnarray}
&&
X(\beta_a,\beta_b,\beta_c)=-\frac{\pi^2}{24\beta_c}
\nonumber \\[-1mm]
&&\label{eq42i} \\[-1mm]
&&~~
+\frac{1}{4}\left(\frac{1}{\beta_a}+\frac{1}{\beta_b}\right)
\int_{0}^{\infty}\!\!dy
\ln\left(1-{\rm e}^{-y}\right)
\nonumber \\[-1mm]
&&\nonumber \\[-1mm]
&&~~
-\frac{1}{2}\left(\frac{1}{\beta_a\beta_c}+\frac{1}{\beta_b\beta_c}
+\frac{1}{\beta_a\beta_b}\right)
\nonumber \\[-1mm]
&&\nonumber \\[-1mm]
&&~~~~~~~\times
\int_{0}^{\infty}\!\!dy\int_{0}^{\infty}\!\!dv
\ln\left(1-{\rm e}^{-\sqrt{y^2+v^2}}
\right)
\nonumber \\[-1mm]
&&\nonumber \\[-1mm]
&&~~
+\frac{1}{\beta_a\beta_b\beta_c}\int_{0}^{\infty}\!\!dy
\int_{0}^{\infty}\!\!dv\int_{0}^{\infty}\!\!dw
\ln\left(1-{\rm e}^{-\sqrt{y^2+v^2+w^2}}\right)
\nonumber \\[-1mm]
&&\nonumber \\[-1mm]
&&~~
+O(\ln\beta_a,\ln\beta_b,\ln\beta_c).
\nonumber
\end{eqnarray}
\noindent
Calculating all integrals and using (\ref{eq42a}) and notations
(\ref{eq42b}) we arrive at the asymptotic expression for the 
nonrenormalized thermal
correction at high temperatures (large separations)
\begin{eqnarray}
&&
\Delta_T{\cal F}_0(a,b,c,T)=-\frac{\pi}{24}(k_BT)^2(a+b+c)
\label{eq42j} \\
&&~~
+
\frac{\zeta(3)}{4\pi}(ac+bc+ab)(k_BT)^3
-\frac{\pi^2}{90}(k_BT)^4abc
\nonumber \\
&&~~
+
O(k_BT\ln\beta_a,k_BT\ln\beta_b,k_BT\ln\beta_c).
\nonumber
\end{eqnarray}

Thus, we have demonstrated that at high temperatures (large separations)
the nonrenormalized
thermal correction really contains the terms of form (\ref{eq29a}) and
(\ref{eq29b}). 
{}From the comparison of (\ref{eq42j}) with (\ref{eq29b}), it follows
\begin{equation}
\alpha_1=\frac{\zeta(3)}{4\pi}(ac+bc+ab),
\quad
\alpha_2=-\frac{\pi}{24}(a+b+c),
\label{eq42k}
\end{equation}
\noindent
i.e., $\alpha_1$ is proportional to the total area of box surface, and 
$\alpha_2$ to the sum of the sides.  These geometrical objects
(together with a box volume) are usually renormalized when replacing
$E_{0,{\rm eff}}$ with $E_0^{\rm ren}$  \cite{2,10,14,15}.
Thus the subtraction of the last three terms on the right-hand side
of (\ref{8e86}) can be interpreted as an additional finite renormalization
giving a physically meaningful temperature-dependent
contribution to the Casimir energy.

As mentioned in Sect.~2, $\alpha_1$ and $\alpha_2$ can be expressed in
terms of the heat kernel coefficients. Thus, keeping in mind that
the heat kernel coefficient $a_{1/2}=-\sqrt{\pi}S/2$ \cite{Vas},
where $S$ is the area of box surface, and comparing (\ref{eq42j}) with
the general asymptotic expression for the free energy \cite{28a},
one arrives at
\begin{equation}
\alpha_1=-\frac{\zeta(3)}{4\pi^{3/2}}a_{1/2}=
\frac{\zeta(3)}{8\pi}S
\label{eq55a}
\end{equation}
\noindent
in agreement with (\ref{eq42k}). Similar comparison of the
general expression \cite{28a} with (\ref{eq42j}) leads to
$\alpha_2=-a_1/24$. The heat kernel coefficient $a_1$ can be
calculated from the known expression for the heat kernel coefficient
of an angle $\theta$ \cite{Nes}
\begin{equation}
c_1(\theta)=\frac{\pi^2-\theta^2}{6\theta}.
\label{eq55b}
\end{equation}
\noindent
In the case of rectangular box $\theta=\pi/2$ and $c_1=\pi/4$.
In the 3-dimensional case this should be multiplied by the length
of all sides leading to
\begin{equation}a_1=4c_1(a+b+c)=\pi(a+b+c).
\label{eq55c}
\end{equation}
\noindent
{}From this, the expression for $\alpha_2$ in (\ref{eq42k})
is reobtained.

The Casimir force
acting between the opposite faces of the box is obtained from 
(\ref{8e86}) and (\ref{eq42k})
\begin{eqnarray}
&&
F_x(a,b,c,T)=-\frac{\partial{\cal F}^{\rm phys}(a,b,c,T)}{\partial a}
=F_x(a,b,c)
\label{8e87}\\
&&~~
+\frac{\pi^2}{a^3}
\sum_{n,l,p=1}^{\infty}\frac{n^2}{\omega_{nlp}
\bigl({\rm e}^{\beta\omega_{nlp}}-1\bigr)}
-\frac{\pi^2(k_BT)^4}{90}bc
\nonumber \\
&&~~
+
\frac{\zeta(3)}{4\pi}(k_BT)^3(b+c)-\frac{\pi}{24}(k_BT)^2.
\nonumber
\end{eqnarray}
\noindent
It is well known \cite{10,31} that
the scalar Casimir energy
$E_0^{\rm ren}(a,b,c)$ is negative and the respective force
$F_x(a,b,c)=-\partial E(a,b,c)/\partial a$ is attractive
for any ratio of $a,\>b$ and $c$. Because of
this, we restrict ourselves to the consideration of a cube
$a=b=c$.

In the limit of low temperature $T\ll T_{\rm eff}$ the leading
terms in (\ref{8e86}) are
\begin{eqnarray}
&&
{\cal F}^{\rm phys}(a,T)=E_0^{\rm ren}(a)-k_BT
\exp\bigl(-\frac{\pi\sqrt{3}}{ak_BT}
\bigr)
\label{8e88}\\
&&~~
+\frac{\pi^2(k_BT)^4}{90}a^3
-\frac{3\zeta(3)}{4\pi}(k_BT)^3a^2+\frac{\pi}{24}(k_BT)^2a,
\nonumber
\end{eqnarray}
\noindent
where the scalar Casimir energy for a cube can be calculated
numerically using (\ref{8e42}) and (\ref{8e11})  with the result
\begin{equation}
E_0^{\rm ren}(a)=-\frac{0.0102}{a}.
\label{8e89}
\end{equation}

At arbitrary temperature it is convenient to represent both the
free energy and force in terms of the dimensionless variable
$t$
\begin{eqnarray}
&&
{\cal F}^{\rm phys}(a,T)=E_0^{\rm ren}(a)
\nonumber \\
&&~~
+\frac{1}{2at}
\sum_{n,l,p=1}^{\infty}\ln\bigl(1-{\rm e}^{-2\pi t\sqrt{n^2+l^2+p^2}}
\bigr)
\nonumber \\
&&~~
+\frac{\pi^2}{1440at^4}
-\frac{3\zeta(3)}{32\pi}\,\frac{1}{at^3}+\frac{\pi}{32}\,\frac{1}{at^2},
\label{8e90} \\
&&
\nonumber \\[-1mm]
&&
\nonumber \\[-1mm]
&&
F_x(a,T)=F_x(a)
\nonumber \\
&&~~
+\frac{\pi}{a^2}
\sum_{n,l,p=1}^{\infty}\frac{n^2}{\sqrt{n^2+l^2+p^2}}
\frac{1}{{\rm e}^{2\pi t\sqrt{n^2+l^2+p^2}}-1}
\nonumber \\
&&~~
-\frac{\pi^2}{1440a^2t^4}
+\frac{\zeta(3)}{16\pi}\,\frac{1}{a^2t^3}-\frac{\pi}{96}\,\frac{1}{a^2t^2},
\nonumber
\end{eqnarray}
\noindent
where the force at zero temperature in accordance with
(\ref{8e89}) is given by
\begin{equation}
F_x(a)=-\frac{0.0102}{3a^2}.
\label{8e91}
\end{equation}

%%%%%%%%%%%%%%%%%%%%%%%%%%%%%%%%%%%%%%%%%%%%%%%%%%%%%%%%%%%%
In Fig.~1(a) we plot the scalar Casimir free energy in a cube
from (\ref{8e90}) as a function of the side length $a$ at $T=300\,$K
(solid line).
As is seen from this figure, the free energy increases
monotonously with the increase of $a$. At large separations 
(not shown in the figure) it approaches to
a constant.
In the same figure the Casimir energy 
$E_0^{\rm ren}=E_0^{\rm ren}(a)$ at zero
temperature is shown by the dashed line. 
In Fig.\ 1(b) the scalar Casimir free energy is plotted as a function
of temperature in the cube with $a=2\,\mu$m. It is seen that at large
temperatures ${\cal F}^{\rm phys}$ is proportional to the temperature
in accordance with the classical limit.

The magnitude of the scalar Casimir force from (\ref{8e90})
(in a logarithmic scale) as a
function of $a$ at $T=300\,$K is shown in Fig.\ 2(a) by
the solid  line. The force is attractive for cubes with
any side length and  its magnitude  goes to zero with the increase of 
$a$. In the same
figure the dashed line shows the magnitude of the Casimir force
(in a logarithmic scale) acting on the opposite faces of the cube
at $T=0$. Fig\ 2(b) shows the Casimir force as a function of 
temperature for the cube with $a=2\,\mu$m. It is seen that
for both negative and positive values of the
free energy the respective Casimir force is attractive.

\section{Electromagnetic Casimir effect in rectangular boxes}

Now we consider the electromagnetic thermal Casimir effect in a
rectangular box with the side lengths $a,\>b$ and $c$.
For the electromagnetic field the renormalized free energy
(\ref{eq29c}), (\ref{5F7}) is specified as
\begin{eqnarray}
&&
{\cal F}_{\rm el}^{\rm phys}(a,b,c,T)=E_{0,{\rm el}}^{\rm ren}(a,b,c)
\label{8e92}\\
&&~~
+k_BT\left[
\sum_{l,p=1}^{\infty}\ln\bigl(1-{\rm e}^{-\beta\omega_{lp}}\bigr)
+\sum_{n,l=1}^{\infty}\ln\bigl(1-{\rm e}^{-\beta\omega_{nl}}\bigr)
\right.
\nonumber \\
&&~~
+\left.
\sum_{n,p=1}^{\infty}\ln\bigl(1-{\rm e}^{-\beta\omega_{np}}\bigr)
+2\sum_{n,l,p=1}^{\infty}\ln\bigl(1-{\rm e}^{-\beta\omega_{nlp}}\bigr)
\right]
\nonumber \\
&&~~
+\frac{\pi^2(k_BT)^4}{45}abc
-\alpha_1^{\rm el}(k_BT)^3-\alpha_2^{\rm el}(k_BT)^2.
\nonumber
\end{eqnarray}
\noindent
Here, $\omega_{nlp}$ is defined in (\ref{8e86a}),
$\omega_{nl}=\omega_{nl0}$, and $\alpha_1^{\rm el}$, $\alpha_2^{\rm el}$ 
have to be determined.
The electromagnetic Casimir energy at $T=0$,
$E_0^{\rm ren}(a,b,c)$, is given by
\begin{eqnarray}
&&
E_{0,{\rm el}}^{\rm ren}(a,b,c)=
-\frac{\pi^2bc}{720a^3}-\frac{\zeta(3)c}{16\pi b^2}+
\frac{\pi}{48}\left(\frac{1}{a}+\frac{1}{b}\right)
\nonumber \\
&&
\phantom{E(a,b,c)}+\frac{\pi}{b}G\left(\frac{c}{b}\right)-
\frac{2}{a}R\left(\frac{b}{a},\frac{c}{a}\right),
\label{8e57}
\end{eqnarray}
\noindent
where the functions $G$ and $R$ are defined in (\ref{8e11}).

In order to find the coefficients $\alpha_1^{\rm el}$ and
$\alpha_2^{\rm el}$, one should determine the asymptotic behavior of
$\Delta_T{\cal F}_{0,\rm el}$ in the limit of high temperatures (large
separations).
In the electromagnetic case the 
nonrenormalized thermal correction can be identically 
rearranged to the form
\begin{eqnarray}
&&
\Delta_T{\cal F}_{0,\rm el}(a,b,c,T)=k_BTY(\beta_a,\beta_b,\beta_c),
\label{eq49a} \\
&&
Y(\beta_a,\beta_b,\beta_c)=2X(\beta_a,\beta_b,\beta_c)
\nonumber \\
&&~~~
+
\sum_{l,p=1}^{\infty}\ln\left(1-{\rm e}^{-\sqrt{\beta_b^2l^2+\beta_c^2p^2}}
\right)
\nonumber \\
&&~~~
+\sum_{n,l=1}^{\infty}\ln\left(1-{\rm e}^{-\sqrt{\beta_a^2n^2+\beta_b^2l^2}}
\right)
\nonumber \\
&&~~~
+\sum_{n,p=1}^{\infty}\ln\left(1-{\rm e}^{-\sqrt{\beta_a^2n^2+\beta_c^2p^2}}
\right),
\nonumber
\end{eqnarray}
\noindent
where the asymptotic behavior of $X(\beta_a,\beta_b,\beta_c)$ at 
small  $\beta_a,\>\beta_b$ and $\beta_c$ was already determined 
in Sect.~4. Taking into account (\ref{eq42a}), (\ref{eq42b}) and
(\ref{eq42j}), it is given by
\begin{eqnarray}
&&
X(\beta_a,\beta_b,\beta_c)=-\frac{\pi^2}{24}\left(\frac{1}{\beta_a}+
\frac{1}{\beta_b}+\frac{1}{\beta_c}\right)
\nonumber \\
&&~~~
+
\frac{\pi\zeta(3)}{4}\left(\frac{1}{\beta_a\beta_b}+
\frac{1}{\beta_a\beta_c}+\frac{1}{\beta_b\beta_c}\right)
\nonumber \\
&&~~~
-\frac{\pi^5}{90}\,\frac{1}{\beta_a\beta_b\beta_c}+
O(\ln\beta_a,\ln\beta_b,\ln\beta_c).
\label{eq49b}
\end{eqnarray}
\noindent
The asymptotic behavior of $Y(\beta_a,\beta_b,\beta_c)$ at small
$\beta_a,\>\beta_b,\>\beta_c$ is obtained in perfect analogy with the
case of the scalar field by the repeated application of the Abel-Plana
formula (\ref{eq42c}) to the remaining three summations in 
(\ref{eq49a}). The result obtained with account of (\ref{eq49b})
is
\begin{eqnarray}
&&
Y(\beta_a,\beta_b,\beta_c)=\frac{\pi^2}{12}\left(\frac{1}{\beta_a}+
\frac{1}{\beta_b}+\frac{1}{\beta_c}\right)
\label{eq49c} \\
&&~~~
-
\frac{\pi^5}{45}\,\frac{1}{\beta_a\beta_b\beta_c}+
O(\ln\beta_a,\ln\beta_b,\ln\beta_c).
\nonumber
\end{eqnarray}
\noindent
Substituting this into (\ref{eq49a}) and using notations (\ref{eq42b}),
one obtains the asymptotic expression for the thermal correction at
high temperatures (large separations)
\begin{eqnarray}
&&
\Delta_T{\cal F}_{0,\rm el}(a,b,c,T)=\frac{\pi}{12}(k_BT)^2(a+b+c)
\label{eq49d} \\
&&~~~
-
\frac{\pi^2}{45}(k_BT)^4abc+
O(k_BT\ln\beta_a,k_BT\ln\beta_b,k_BT\ln\beta_c).
\nonumber
\end{eqnarray}

It is notable that (\ref{eq49d}) does not contain a contribution
proportional to the surface area of the box [compare with (\ref{eq42j})
for the scalar field]. In the electromagnetic case such a contribution is
absent also in the divergent Casimir energy of the box $E_{0,{\rm el}}$
at zero temperature \cite{2,10,14,15}. 
Thus, from (\ref{eq49d}) and (\ref{eq29b}) one arrives at
\begin{equation}
\alpha_1^{\rm el}=0, \quad
\alpha_2^{\rm el}=\frac{\pi}{12}(a+b+c).
\label{eq49e}
\end{equation}

 The Casimir force acting between
the opposite faces of a box is obtained as the negative
derivative of (\ref{8e92}) with respect to $a$,
\begin{eqnarray}
&&
F_{x,{\rm el}}(a,b,c,T)=F_{x,{\rm el}}(a,b,c)+\frac{\pi^2}{a^3}\left[
\sum_{n,l=1}^{\infty}\frac{n^2}{\omega_{nl}
\bigl({\rm e}^{\beta\omega_{nl}}-1\bigr)}\right.
\nonumber \\
&&~~
\left.+
\sum_{n,p=1}^{\infty}\frac{n^2}{\omega_{np}
\bigl({\rm e}^{\beta\omega_{np}}-1\bigr)}
+
2\sum_{n,l,p=1}^{\infty}\frac{n^2}{\omega_{nlp}
\bigl({\rm e}^{\beta\omega_{nlp}}-1\bigr)}\right]
\nonumber \\
&&~~
-
\frac{\pi^2(k_BT)^4}{45}bc+\frac{\pi}{12}(k_BT)^2.
\label{8e93}
\end{eqnarray}

It has been known that the electromagnetic
Casimir energy inside a box at $T=0$, $E_{0,{\rm el}}^{\rm ren}(a,b,c)$,
can be both positive and negative and the Casimir force,
$F_{x,{\rm el}}(a,b,c)=-\partial E_{0,{\rm el}}^{\rm ren}(a,b,c)/\partial a$,
can be both attractive and repulsive depending on
the ratio of the sides $a,\>b$ and $c$ \cite{2,10}.
Here
we consider in more detail the thermal electromagnetic
Casimir effect for a cube $a=b=c$ where the electromagnetic
Casimir energy at zero temperature computed by (\ref{8e57})
is positive,
\begin{equation}
E_{0,{\rm el}}^{\rm ren}(a)=\frac{0.09166}{a}>0,
\label{8e93a}
\end{equation}
and the force (\ref{8e93}) is repulsive.

For a cube the
electromagnetic Casimir free energy (\ref{8e92}) and
force (\ref{8e93}) are given by
\begin{eqnarray}
&&
{\cal F}_{\rm el}^{\rm phys}(a,T)=E_{0,{\rm el}}^{\rm ren}(a)+
\frac{3}{2at}\!
\sum_{n,l=1}^{\infty}\ln\bigl(1-{\rm e}^{-2\pi t\sqrt{n^2+l^2}}\bigr)
\nonumber \\
&&\,
+\frac{1}{at}\!\!\sum_{n,l,p=1}^{\infty}\!\!\ln\bigl(1-
{\rm e}^{-2\pi t\sqrt{n^2+l^2+p^2}}\bigr)
+\frac{\pi^2}{720at^4}-\frac{\pi}{16at^2},
\nonumber \\
&&
F_{x,{\rm el}}(a,T)=F_{x,{\rm el}}(a)
\label{8e94}\\
&&~~
+\frac{2\pi}{a^2}\left[
\sum_{n,l=1}^{\infty}\frac{n^2}{\sqrt{n^2+l^2}}\,
\frac{1}{{\rm e}^{2\pi t\sqrt{n^2+l^2}}-1}\right.
\nonumber \\
&&~~
\left.+
\sum_{n,l,p=1}^{\infty}\frac{n^2}{\sqrt{n^2+l^2+p^2}}
\frac{1}{{\rm e}^{2\pi t\sqrt{n^2+l^2+p^2}}-1}\right]
\nonumber \\
&&~~
-
\frac{\pi^2}{720a^2t^4}+\frac{\pi}{48a^2t^2},
\nonumber
\end{eqnarray}
\noindent
where the force at $T=0$ is
\begin{equation}
F_{x,{\rm el}}(a)=\frac{0.09166}{3a^2}.
\label{8e95}
\end{equation}

In Fig.~3(a) we plot the electromagnetic Casimir free energy in
a cube as a function of $a$ at $T=300\,$K (solid line).
In the same figure the Casimir energy at $T=0$ is shown
by the dashed line. 
As is seen in this figure, the electromagnetic Casimir free energy
decreases with the increase of separation.
Similar to the scalar case, at large separations 
${\cal F}_{\rm el}^{\rm phys}$ approaches to a constant.
In Fig.~3(b) the electromagnetic Casimir free energy is shown
as a function of temperature for a cube with $a=2\,\mu$m.
The free energy decreases with the increase of $T$.
At high temperatures ${\cal F}_{\rm el}^{\rm phys}$ 
demonstrates the classical limit.
The respective thermal electromagnetic Casimir force at $T=300\,$K,
as a function
of $a$, is shown in Fig.\ 4(a) by the solid line.
It is positive (i.e., repulsive) for cubes with any side lengths.
Thus, thermal effects for cubes in the electromagnetic case increase the
strength of the Casimir repulsion.
 The dashed line in Fig.\ 4(a) shows the
electromagnetic Casimir force at $T=0$ as a function of $a$.
This force is given by (\ref{8e95}), i.e., it is always
repulsive.  
Fig.\ 4(b) demonstrates the electromagnetic Casimir force 
in a cube of side length $a=2\,\mu$m
as a function of temperature. It is seen that the force increases
with increasing temperature.

Note that the obtained results differ from those found in \cite{25},
where the terms of order $(k_BT)^4$ and of lower orders in the Casimir
free energy were obtained in the high-temperature regime. Also, the 
Casimir free energy in \cite{25} both for the scalar and electromagnetic
field is always a decreasing function of temperature on the opposite
to our result in Fig.\ 1(b). This is explained by the fact that in
\cite{25} the subtractions of the contributions from the 
blackbody radiation and of the terms proportional to the box surface
area and to the sum of its sides were not made. 

The above equations
(\ref{8e86}), (\ref{8e87}), (\ref{8e92}) and (\ref{8e93})
can be used to compute the scalar and electromagnetic free energy and
force for boxes with arbitrary side lengths $a,\>b$ and $c$. 
Specifically it follows that the temperature-de\-pen\-dent contribution to the
electromagnetic Casimir force
[which is obtained as $-\partial\Delta_T{\cal F}/\partial a$ from the
physical thermal correction defined in (\ref{eq29c})]
can be both positive and negative depending on the side
lengths $a,\>b$ and $c$.
On the one hand, as shown above (see Fig. 4), for a cube 
$a\times a\times a$ the temperature-dependent contribution to the
Casimir force is positive and computations show that
this is preserved for any box
$a\times b\times b$ with $a>b$. On the other hand, for boxes
with $b=c=10\,\mu$m and $a_1=2.942\,\mu$m or $a_2=34.29\,\mu$m
the Casimir energy at $T=0$ is equal to zero \cite{2,10}. 
Computations using (\ref{8e93}) show that 
for the box $a_1\times b\times b$
the temperature-dependent contribution to the
force is negative, whereas for the box $a_2\times b\times b$ it is positive.

The thermal correction to the Casimir 
energy and force acting on a piston
were investigated \cite{19} for the scalar
field with Dirichlet or Neumann boundary conditions using (25). 
The electromagnetic
Casimir free energy and
force acting on a piston were found in the case
of ideal metal rectangular boxes and cavities with general cross
section \cite{19}. 
In the limit of low temperatures the thermal correction 
to the Casimir force on a piston was shown to be exponentially small.
In the limit of medium temperature \cite{19}
obtains the term of order $(k_BT)^4$ and of lower 
orders in the electromagnetic Casimir free energy (see
(44) and (47) in \cite{19}). 
This results in the contribution to the force acting on a piston which
 increases with the increase of the
temperature, depends on $\hbar$ and $c$ and does not depend on the
position of a piston. 
The same results are obtained from the application of (\ref{eq29c}).
This is because the contribution of the blackbody radiation to the energy
of entire box is equal to $-abcf_{\rm bb}$ and does not depend on the position
of a piston. The contributions from the terms proportional to the surface
areas and the sums of the side lengths for the two boxes to the left and to the
right of a piston are also independent of the piston position.

\section{Conclusions and discussion}

In the above we have reconsidered the old problem of thermal Casimir
force in rectangular boxes made of ideal metal. Our main concern was with
the appropriate definition of the Casimir free energy inside some
closed volume. We have stressed that such a definition must include
a subtraction of the energy of blackbody radiation confined in this
volume and of two other terms proportional to $(k_BT)^3$ and  $(k_BT)^2$
which are contained in the high temperature (large separation) 
asymptotic expression for the thermal correction to the Casimir
energy $\Delta_T{\cal F}_0$ in (\ref{5F7}).
If this is done, the thermal Casimir force turns into zero when
the characteristic sizes of the system go to infinity.
According to the results of Sect. 3, the suggested expression
(\ref{eq29c}) for the Casimir free energy
is applicable to the limiting case of two plane parallel
ideal metal planes and leads to the same result as obtained by many
authors in the framework of thermal quantum field theory in Matsubara
formulation. Both the free energy and the thermal Casimir force were
calculated for the massless scalar field and for the electromagnetic
field inside ideal metal rectangular boxes. 
For the scalar field with
Dirichlet boundary conditions the thermal
Casimir force is always attractive. 
It was shown that
for the electromagnetic field in rectangular boxes 
the temperature-dependent contribution to
Casimir force can be both positive and negative depending on
the side lengths.
 At high temperatures the Casimir free energy
is proportional to $k_BT$, i.e., the classical limit is achieved, like
it holds for two plane parallel planes. The suggested expression for the
Casimir free energy leads to the cancellation of all terms of order
$(k_BT)^4$,  $(k_BT)^3$ and $(k_BT)^2$
at high temperatures (large separations). 
This is in accordance with what the physical
intuition suggests because such terms depend on the Planck constant and,
thus, are of quantum origin.

The above results are obtained for rectangular boxes with
the Dirichlet boundary conditions (scalar case) or for ideal
metal boxes (electromagnetic case). We would like to emphasize that the
final resolution of the problems of Casimir repulsion and of the
thermal Casimir effect in rectangular boxes requires the inclusion of
real material properties of the boundary surfaces. In the case of
closed quantization volumes this is a challenging issue which remains
to be explored.

\section*{Acknoledgments}

The authors thank M.~Bordag for numerous helpful discussions.
G.L.K.\ and V.M.M.\ are grateful to the Center of Theoretical Studies
and the Institute for Theoretical Physics, Leipzig University, for their
kind hospitality.
This work was supported by Deutsche Forschungsgemeinschaft,
 Grant No 436 RUS 113/789/0--4.

%%%%%%%%%%%%%%%%%%%%%%%%%%%%%%%%%%%

%%%%%%%%%%%%%
%\end{document}
%%%%%%%%%%%%%%%%%%%%%%%%%%%%%%%%%%%%%%%%%%%%%%%
\begin{figure*}[h]
\vspace*{1cm}
\centerline{
\includegraphics{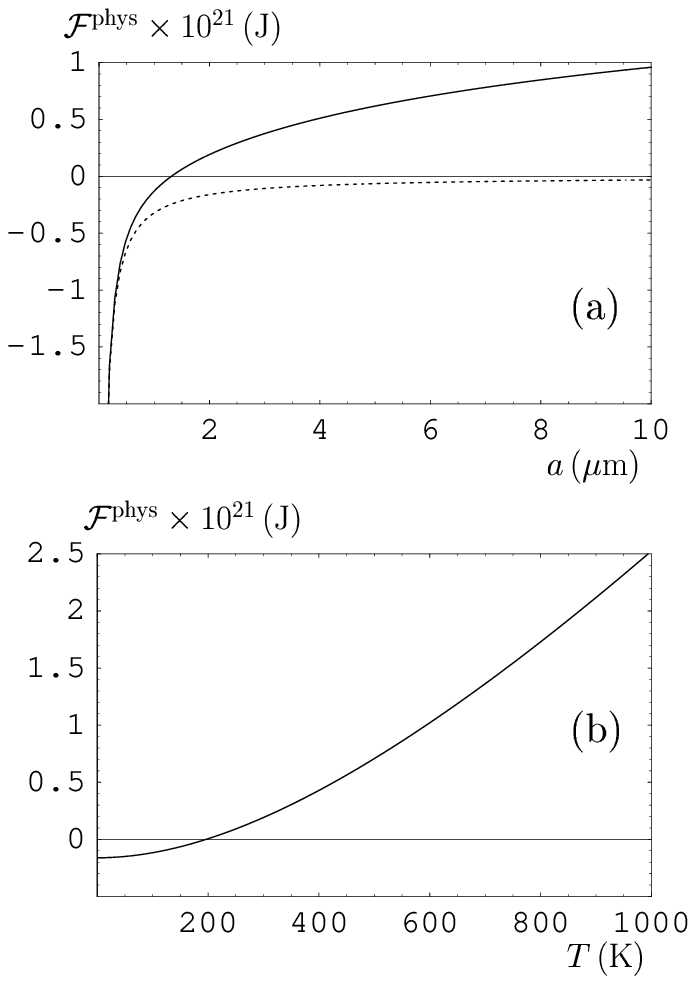}
}
\vspace*{-17cm}
\caption{The scalar Casimir free energy
for a cube as a function of (a) a side length $a$ at $T=300\,$K 
(solid line; the dashed line shows the energy at $T=0$) and (b)
temperature at $a=2\,\mu$m.
}
\end{figure*}
%%%%%%%%%%%%%%%%%%%%%%%%%%%%%%%%%%%%%%%%%%%%%%
%%%%%%%%%%%%%%%%%%%%%%%%%%%%%%%%%%%%%%%%%%%%%%%
\begin{figure*}[h]
\vspace*{1cm}
\centerline{
\includegraphics{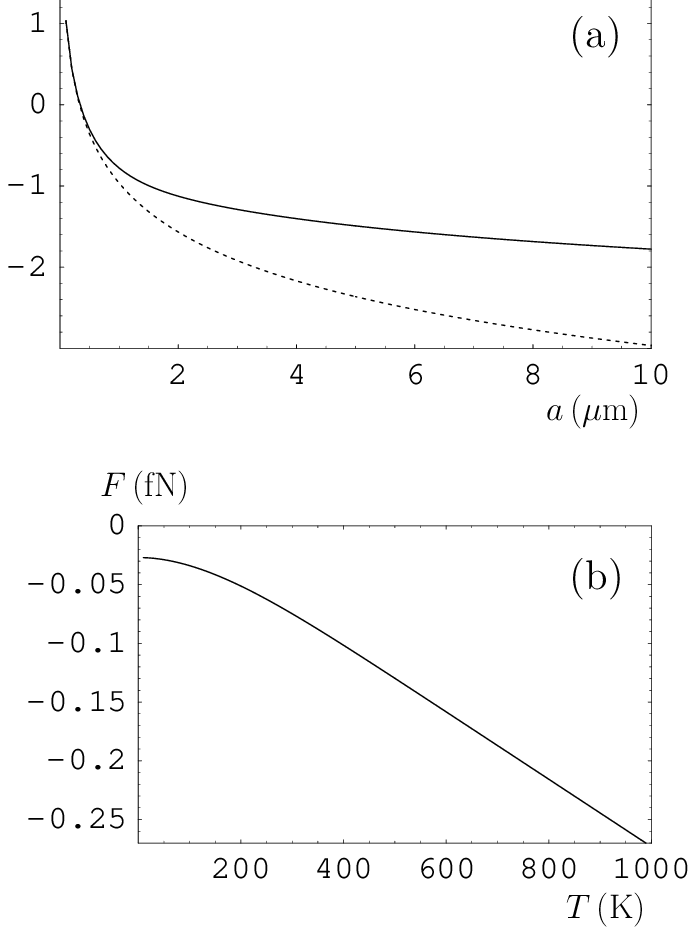}
}
\vspace*{-17cm}
\caption{The scalar Casimir force between
the opposite
faces of a cube as a function of (a) a side length $a$ at $T=300\,$K 
(solid line; the dashed line shows the force at $T=0$) and (b)
temperature at $a=2\,\mu$m.
}
\end{figure*}
%%%%%%%%%%%%%%%%%%%%%%%%%%%%%%%%%%%%%%%%%%%%%%
%%%%%%%%%%%%%%%%%%%%%%%%%%%%%%%%%%%%%%%%%%%%%%%
\begin{figure*}[h]
\vspace*{1cm}
\centerline{
\includegraphics{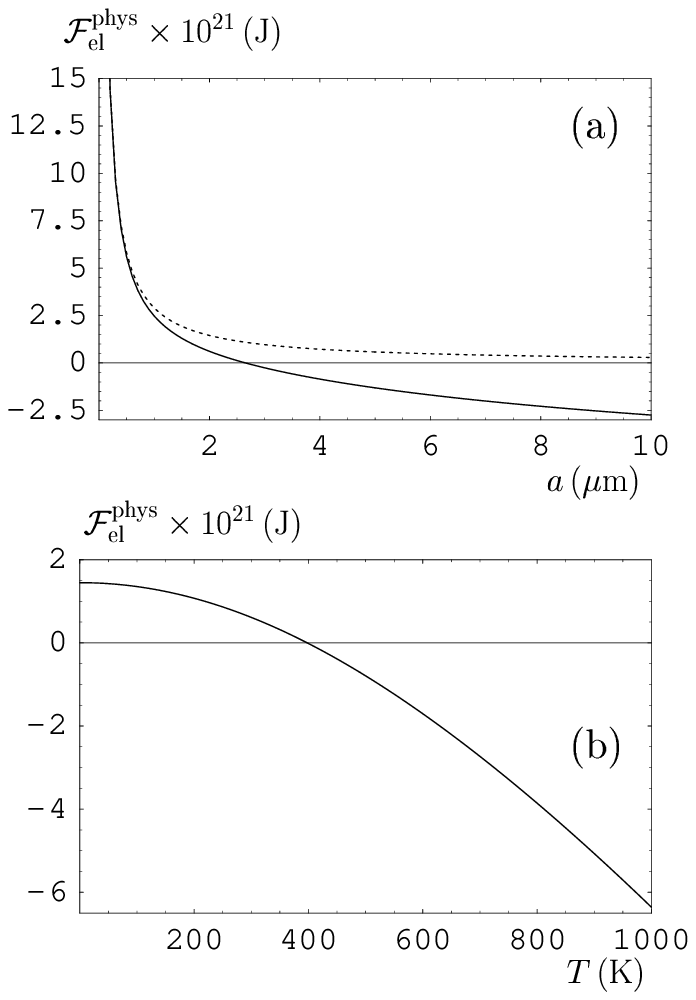}
}
\vspace*{-17cm}
\caption{The electromagnetic Casimir free energy for
 a cube as a function of (a) a side length $a$ at $T=300\,$K 
(solid line; the dashed line shows the energy at $T=0$) and (b)
temperature at $a=2\,\mu$m.
}
\end{figure*}
%%%%%%%%%%%%%%%%%%%%%%%%%%%%%%%%%%%%%%%%%%%%%%
%%%%%%%%%%%%%%%%%%%%%%%%%%%%%%%%%%%%%%%%%%%%%%%
\begin{figure*}[h]
\vspace*{1cm}
\centerline{
\includegraphics{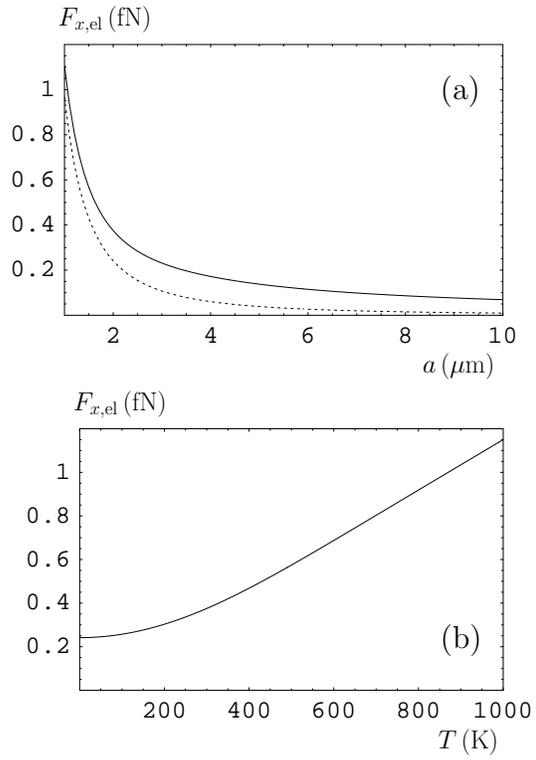}
}
\vspace*{-17cm}
\caption{The electromagnetic Casimir force between
the opposite
faces of a cube as a function of (a) a side length $a$ at $T=300\,$K 
(solid line; the dashed line shows the force at $T=0$) and (b)
temperature at $a=2\,\mu$m.
}
\end{figure*}
%%%%%%%%%%%%%%%%%%%%%%%%%%%%%%%%%%%%%%%%%%%%%%
\end{document}